\documentclass[prd,twocolumn,showpacs,preprintnumbers,amsmath,amssymb]{revtex4}

\usepackage{graphicx}
\usepackage{dcolumn}
\usepackage{bm}
\usepackage{wrapfig} 
\usepackage{multirow}
\usepackage{color}

\newcommand{\beq}{\begin{eqnarray}} 
\newcommand{\eeq}{\end{eqnarray}}

\begin{document}

\preprint{CERN--PH--TH/2013--177}\vspace*{0.5cm}

\title{Supersymmetry with Light Dark Matter \\ 
confronting the recent CDMS and LHC Results}

\author{Alexandre Arbey}
\email{alexandre.arbey@ens-lyon.fr}
\affiliation{Centre de Recherche Astrophysique de Lyon, Observatoire de Lyon,\\
\mbox{Saint-Genis Laval Cedex, F-69561, France; CNRS, UMR 5574;}\\ 
Ecole Normale Sup\'erieure de Lyon, France;\\
\mbox{Universit\'e de Lyon, Universit\'e Lyon 1, F-69622~Villeurbanne Cedex, France}\\
\mbox{and CERN Theory Division, CH-1211 Geneva, Switzerland}}%

\author{Marco Battaglia}
\email{MBattaglia@lbl.gov}
\affiliation{University of California at Santa Cruz, \\
Santa Cruz Institute of Particle Physics, CA 95064, USA\\
and CERN, CH-1211 Geneva, Switzerland}%

\author{Farvah Mahmoudi}%
\email{mahmoudi@in2p3.fr}
\affiliation{Clermont Universit\'e, Universit\'e Blaise Pascal, CNRS/IN2P3,\\
LPC, BP 10448, F-63000 Clermont-Ferrand, France\\
and CERN Theory Division, CH-1211 Geneva, Switzerland}%


\begin{abstract}
\vspace*{0.5cm}
We revisit MSSM scenarios with light neutralino as a dark matter candidate in view of the latest LHC and dark matter direct and indirect detection experiments. We show that scenarios with a very light neutralino ($\sim 10$ GeV) and a scalar bottom quark close in mass, can satisfy all the available constraints from LEP, Tevatron, LHC, flavour and low energy experiments and provide solutions in agreement with the bulk of dark matter direct detection experiments, and in particular with the recent CDMS results. 
\end{abstract}

\maketitle

\section{Introduction}

Direct searches for dark matter (DM) at underground experiments and for Higgs bosons and new particles at the CERN LHC collider represent powerful probes 
into extensions of the Standard Model (SM), including a weakly interacting massive particle (WIMP) protected by a symmetry ensuring its stability. 

In the last few years, the DAMA/LIBRA~\cite{Bernabei:2010mq}, CoGeNT~\cite{Aalseth:2011wp} and CRESST-II~\cite{Angloher:2011uu} experiments have all reported excesses of events over their estimated backgrounds, which can be interpreted as due to low-mass dark matter WIMPs interacting in their detectors. These claims had to be confronted to the negative results of searches conducted by the CDMS~\cite{Ahmed:2010wy} and XENON~\cite{Aprile:2012nq} experiments, as well as the absence of new physics reported by the ATLAS and CMS experiments at the LHC. In an earlier paper~\cite{Arbey:2012na}, we showed that the events of DAMA/LIBRA, CoGeNT and CRESST-II were consistent with a Supersymmetric (SUSY) scenario with a light neutralino, $\tilde{\chi}^0_1$, and an almost
degenerate scalar lepton, gaugino, or scalar bottom,  which would have escaped searches based on hadronic jets plus missing transverse energy (MET) signatures 
at the LHC, due to the very low transverse energy of the jets. The scenario with scalar bottom, $\tilde{b}_1$, as the next lightest SUSY particle (NLSP) 
gave the best in agreement with the available experimental results and with a neutralino mass below $\sim$20~GeV. Here we pursue further this scenario. 
Other recent studies on neutralino DM in Supersymmetry have been presented 
in~\cite{Grothaus:2012js,Boehm:2013qva,Choudhury:2013jpa,Fowlie:2013oua,Calibbi:2013poa,Han:2013gba}. 

The recent analysis of the CDMS-II data has isolated three possible signal events, with a small expected background~\cite{Agnese:2013rvf}. If these events are due to the interaction of WIMPs in the CDMS detector, the WIMP mass and scattering cross section would be comparable to those already highlighted by the other experiments reporting possible excesses of events. The first data from the PLANCK satellite~\cite{Ade:2013zuv} have improved the determination of the DM relic density, $\Omega_{\mathrm{CDM}} h^2$, through the study of the Cosmic Microwave Background (CMB). The ATLAS and CMS experiments at LHC have discovered a Higgs-like scalar particle with a mass of $\simeq$126~GeV and significantly extended the constraints on SUSY with analyses sensitive also to the production of weakly interacting particle partners. The Higgs boson discovery and the first determination of the rates of its decays are crucial to the interpretation of DM results. The Higgs boson participates to the processes of DM annihilation and WIMP scattering on nucleons, and new massive particles, such as WIMPS, should couple to the Higgs field, thus affecting the Higgs decay pattern. 

This paper discusses the viability of a SUSY interpretation of the reported excesses in DM direct searches, all pointing to a WIMP with mass in the range 5 -- 25~GeV and scattering cross section, $\sigma^{\mathrm{SI}}_{\chi p} \simeq$ 10$^{-6}$ -- 10$^{-4}$~pb, in the light of the latest LHC results. We consider the specific light, almost degenerate neutralino-sbottom scenario identified in~\cite{Arbey:2012na}. In this scenario, a large mixing angle $\theta_b$ corresponding to a mainly right-handed $\tilde{b}_1$ decouples the light scalar bottom from the $Z$ boson and makes possible to have a very light $\tilde{b}_1$ if $\tilde{b}_R$ is light. Compared to our previous study,  here we focus on the interplay between SUSY low mass particle partners and the Higgs boson signal strengths and that between the Higgs mass and the WIMP annihilation and WIMP-nucleon scattering processes, relevant to the relic DM density and  direct and indirect detection signals, in this specific scenario. The first determinations of the $\sim$126~GeV Higgs-like scalar particle by the LHC experiments with a useful accuracy and the results of a broad spectrum of searches have significant implications on SUSY scenarios with very light particles, which are now discussed in detail. In this study, we consider these constraints on the proposed light, almost degenerate neutralino-sbottom scenario coming from electroweak precision data, LHC SUSY and exotic searches and Higgs data, simulating events for the selected MSSM points and explicitly checking that these are not excluded by the latest, preliminary 8~TeV LHC results. We also discuss the results obtained using different approaches for the computation of the WIMP scattering cross section, compared to that used in~\cite{Arbey:2012na}, in response to a recent study~\cite{Gondolo:2013wwa}.

This paper is organised as follows. In Section II we review the supersymmetric dark matter in view of the available constraints. 
In Section III we describe in detail the LHC constraints. The pMSSM scenario with a light neutralino and almost degenerate scalar 
bottom $\tilde{b}_1$ is discussed in Section IV, while Section V gives our conclusions.

\section{Supersymmetric Light Dark Matter}

In this study we consider the minimal supersymmetric extension of the SM (MSSM) with the $\tilde{\chi}^0_1$ as the lightest supersymmetric particle (LSP) and R-parity conservation. We take the SUSY mass terms and trilinear couplings as independent free parameters, leading to the 19-parameter phenomenological MSSM (pMSSM) model. If the $\tilde{\chi}^0_1$ mass is as light as the WIMP particle compatible with the DAMA/LIBRA, CoGeNT, CRESST and CDMS results, appropriate mechanisms must be in place not to exceed the $\Omega_{\mathrm{CDM}} h^2$ upper bound derived by the WMAP~\cite{Hinshaw:2012aka} and PLANCK~\cite{Ade:2013zuv} CMB data and also to escape collider precision data, such as the tight constraints from the $Z$ lineshape measurements at LEP. 
In our earlier study~\cite{Arbey:2012na}, we could identify only one viable scenario in the pMSSM consistent with the $\Omega_{\mathrm{CDM}} h^2$ upper bound, the excesses of events in the direct detection experiments and the LEP constraints. In this scenario the $\tilde{b}_1$ scalar quark is very light and the squark mixing angle $\theta_b$ large, close to $\pi/2$, to make the $\tilde{b}_1$, then mainly $\tilde{b}_R$, almost degenerate with the $\tilde{\chi}^0_1$ LSP and with reduced coupling to the $Z$. The concurrent light $\tilde{b}_1$ mass and its decoupling from the $Z$ through a large value of $\theta_b$ offered a compelling MSSM scenario with a low mass WIMP compatible with the available data.

\subsection{Tools}

The tools used to perform the scans and the analysis have been presented in Ref.~\cite{Arbey:2011un,Arbey:2011aa}. 
Most relevant to this study are the calculations of the neutralino scattering cross sections and relic density. 
These are calculated using {\tt DarkSUSY 5.1.1}~\cite{Gondolo:2004sc} and {\tt SuperIso Relic v3.2}~\cite{superiso,superiso_relic}, respectively.
We comment on the comparison of {\tt DarkSUSY} with the results obtained with {\tt micrOMEGAs}~\cite{Belanger:2008sj}, used in our previous study, 
in Section~\ref{sec:direct}. SUSY particle spectra are calculated using {\tt SOFTSUSY 3.2.3}~\cite{softsusy}. {\tt HDECAY (5.10)}~\cite{Djouadi:1997yw} and 
{\tt SDECAY}~\cite{Muhlleitner:2003vg} compute the decay branching fractions for the Higgs and SUSY particles, respectively. 
In order to check the compatibility of selected pMSSM points with various searches at LEP and the LHC we simulate event sample and perform a parametric 
simulation for event reconstruction. Events are generated with {\tt MadGraph 5}~\cite{Alwall:2011uj} and {\tt Pythia 8.150}~\cite{pythia8} and detector 
fast simulation is performed using {\tt Delphes 3.0}~\cite{Ovyn:2009tx}.

\subsection{Electro-weak and $e^+e^-$ Search Constraints}
SUSY searches at LEP and the Tevatron have set stringent constraints on light supersymmetric particle masses. However, their sensitivity depends on the mass splittings of the SUSY particles and the LSP, $\Delta M$. Here, we apply the same mass limits as in our previous study, and comment on the most constraining measurements.

The most constraining LEP observable for this scenario is the $Z$ boson width.  The $Z$ boson decay to two neutralinos contributes to the invisible $Z$ width, measured to be $\Gamma_{\rm inv} = (499.0 \pm 1.5)$ MeV, consistent with the SM prediction~\cite{ALEPH:2005ab}. 
We impose the decay width to two neutralinos to be smaller than 3~MeV, i.e.\ within the 
measurement accuracy. Since in our scenario $\tilde{\chi}^0_1$ is bino-like and couples only very weakly to the $Z$ boson, this constraint is easily satisfied. 
The scalar bottom quark is also very light and the $Z$ boson can decay into a $\tilde{b}_1 \tilde{b}_1$ pair. This decay contributes to the total $Z$ width. The LEP measurements give $\Gamma_{\rm tot} = 
(2495.2 \pm 2.3)$~MeV~\cite{ALEPH:2005ab}. 
We require that the sum of the $Z$ decay widths to neutralino and scalar bottom pairs is smaller than 5~MeV, which corresponds to a 2$\sigma$ deviation from the measured value, accounting for the theoretical uncertainty from the sbottom mixing calculation. Since the sbottom mixing angle, $\theta_b$, is close to $\pi/2$ and the $\tilde{b}_1$ is mainly right-handed, the coupling of the $\tilde{b}_1$ to the $Z$ boson is reduced, leading to a small decay width, so that a significant fraction of the pMSSM points corresponding to our scenario are in agreement with this constraint.

A third important observable is the ratio $R_b$ of the $Z$ decay width to two bottom quarks over the $Z$ total hadronic width. This has been measured very precisely~\cite{ALEPH:2005ab}. The presence of light sbottoms could indeed modify at loop level the effective coupling of the bottoms to the $Z$. We compute $R_b$ for the points passing the $Z$ decay width constraints and find that it agrees within one standard deviation with the experimental measurements as a result of the reduced coupling of the $\tilde{b}_1$ to the $Z$.

Another relevant observable is the forward-backward asymmetry on the $Z$ peak in the $b \bar{b}$ channel \cite{ALEPH:2005ab}, which presents 2.5$\sigma$ discrepancy between the SM and the measured values. In our scenario, while the discrepancy is not improved by the presence of the light sbottoms, our points are in agreement with the experimental result at the 3$\sigma$ level.

\begin{figure}[h!]
\includegraphics[width=0.7\columnwidth]{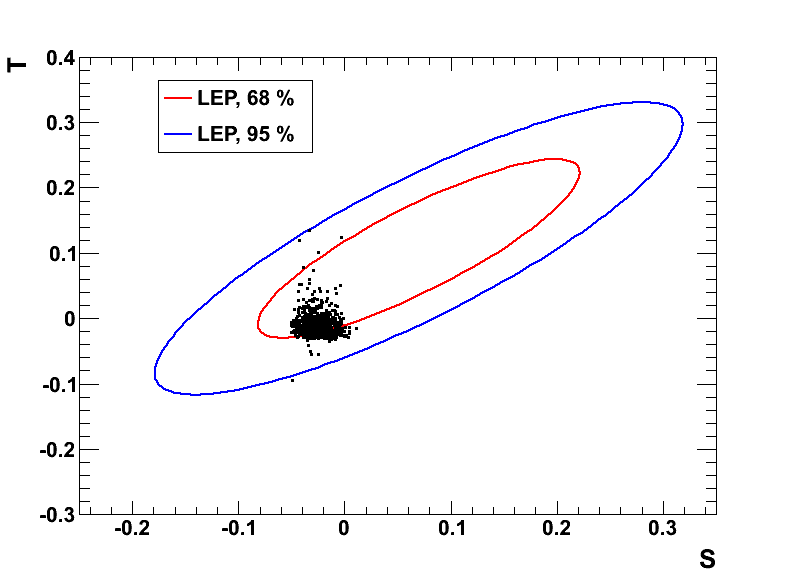}
\vspace*{-0.40cm}
\caption{Valid pMSSM points corresponding to the light neutralino, almost degenerate $\tilde{b}_1$ scenario in the plane $(S,T)$. 
 The ellipses correspond to the LEP allowed regions at 68\% (red) and 95\% (blue) C.L.~\cite{ALEPH:2005ab}.}
\label{fig:stu}
\end{figure}

Constraints from the $S$, $T$ and $U$ parameters~\cite{Kennedy:1988sn,Peskin:1991sw}, encoding the oblique corrections, i.e.\ the radiative corrections to weak processes involving light particles, need also to be considered. In particular, SUSY contributions to these parameters arise also from squark and neutralino loops~\cite{Dobado:1997up,Cho:1999km,Martin:2004id}. The SUSY contributions to $S$, $T$ and $U$ parameters for the points selected in this analysis have been computed and are found to be all compatible with the LEP measurements at 95\% C.L., as shown in Fig.~\ref{fig:stu}. 

Searches for SUSY particles in $e^+e^-$ collisions have been conducted at various energies before LEP. In particular, TRISTAN operated at 52 $< \sqrt{s} <$ 57~GeV, 
where $\tilde{b}_1 \tilde{b}_1$ pairs could be kinematically produced. However, due to the small coupling to the $Z$, the production cross section for 
$e^+e^- \rightarrow \tilde{b}_1 \tilde{b}_1$ is in the range 1.2--0.2~pb for 15 $< M_{\tilde{b}_1} <$ 25~GeV. We estimate the efficiency of the TRISTAN 
analysis~\cite{Adachi:1988fk} by applying its selection cuts on the samples of $e^+e^- \rightarrow \tilde{b}_1 \tilde{b}_1$ generated with 
{\tt Pythia 6.424}~\cite{Sjostrand:2006za} at $\sqrt{s} = $ 57~GeV.  These selection criteria, based on the requirement of a total visible energy in excess of 
10~GeV, small sphericity and two reconstructed jets with large acoplanarity angle, have an efficiency estimated to be 0.03, 0.20 and 0.41 for our pMSSM points 
with a mass splitting, $\Delta M$ = 5, 7 and 10~GeV respectively. This result agrees with the efficiency values reported 
by the experiment~\cite{Adachi:1988fk}, i.e.\ an efficiency in excess to 0.40 for $\Delta M$ = 13~GeV and decreasing towards zero for $\Delta M$ below 8~GeV. 
From these results we can conclude that, with a total data statistics of $\simeq$11~pb$^{-1}$ for TRISTAN, no signal of these events could be obtained for points 
having $\Delta M \le$ 7.  

At LEP-2, the searches for $e^+e^- \rightarrow \tilde{b}_1 \tilde{b}_1$ pair production have excluded scalar bottom quarks up to $\sim$100~GeV, with 
the exception of highly degenerate scenarios. The efficiency of the selection cuts applied in the LEP-2 searches, mostly to reduce 
$\gamma \gamma$ background has been tested on simulated events found to be $\sim$0.15 at $\Delta M$ = 7~GeV and $\le$0.10 at $\Delta M \le$ 5~GeV. 
Since the typical $e^+e^- \rightarrow \tilde{b}_1 \tilde{b}_1$ production cross section at 200~GeV is $\simeq$0.2~pb for 15 $< M_{\tilde{b}_1} <$ 25~GeV, 
this results in a product of signal cross section times efficiency of 0.03~pb and less for $\Delta M \le$ 7 ~GeV, which are therefore not excluded 
by the combined LEP-2 searches. In summary, scalar bottom quarks with 15 $< M_{\tilde{b}_1} <$ 25~GeV, small $\cos \theta_b$ and mass splitting to the 
lightest neutralino $<7$~GeV are not excluded by direct scalar quark searches at $e^+e^-$ colliders.

Finally, the process $e^+e^- \rightarrow \tilde{\chi}^0_1 \tilde{\chi}^0_2$ is suppressed since the lightest neutralino, $\tilde{\chi}^0_1$, is bino-like 
and the second lightest, $\tilde{\chi}^0_2$, is wino-like. In general the $\tilde{\chi}^0_2$ can be chosen to be heavier then 200~GeV, thus ensuring that the 
$\tilde{\chi}^0_1 \tilde{\chi}^0_2$ pairs could not be produced at LEP-2. But the process has a cross section of 
less than 0.1~fb, even when the process is kinematically accessible, as for the case $M_{\tilde{\chi}^0_2}$ = 
150~GeV, due to the coupling suppression.

\subsection{Vacuum Stability}

The MSSM introduces several additional scalars, resulting in a more complex scalar potential. Hence, the stability of the vacuum expectation value (VEV) configurations and the possibility of a tunnelling to other minima of the potential need to be checked. To address this question, which was not considered in~\cite{Arbey:2012na}, we use the program {\tt Vevacious}~\cite{Camargo-Molina:2013qva}, which determines the global minimum of the one loop effective scalar potential for each MSSM point. If the local minimum is global, the vacuum is stable. Otherwise, the program computes the tunnelling time from the local to the global minimum. This should be compared to the age of the Universe, excluding points for which the vacuum is short-lived. About 85\% of the accepted pMSSM points in our scenario have stable vacuum, 5\% have a long-lived vacuum, and 10\% have a short-lived vacuum.

\subsection{Direct Detection}
\label{sec:direct}
The results of direct detection experiments reporting possible excesses of signal-like events, correspond to a light WIMP with large value of the scattering cross section. 

Our pMSSM scenario has a light $\tilde{\chi}^0_1$ and an almost degenerate $\tilde{b}_1$ with a mass splitting 
of order of the bottom mass. 
We observe that the calculation of the cross section for direct detection in such a scenario requires special care. In this specific regime 
the general effective Lagrangian approach is not quite appropriate and requires a special treatment, for example treating the $b$ quark as 
a heavy quark throughout the full calculation, including the twist-2 terms. Applying the default general formula, as used in {\tt micrOMEGAs} 
adopted in our earlier study~\cite{Arbey:2012na}, in the case where $M_{\tilde{b}_1} \approx M_{\tilde{\chi}_1^0} - m_b$, may reveal a spurious 
pole that, erroneously, enhances the scattering cross section. Ref.~\cite{Gondolo:2013wwa} has recently reconsidered the calculation of this 
cross section for the specific case considered here, based on the Drees and Nojiri (DN) treatment~\cite{Drees:1993bu}. The scattering cross 
section obtained using the DN treatment implemented in {\tt DarkSUSY 5.1.1} still provides us with a sizeable amount of pMSSM points consistent 
with CDMS and other data. 

\begin{figure}[h!]
\includegraphics[width=0.65\columnwidth]{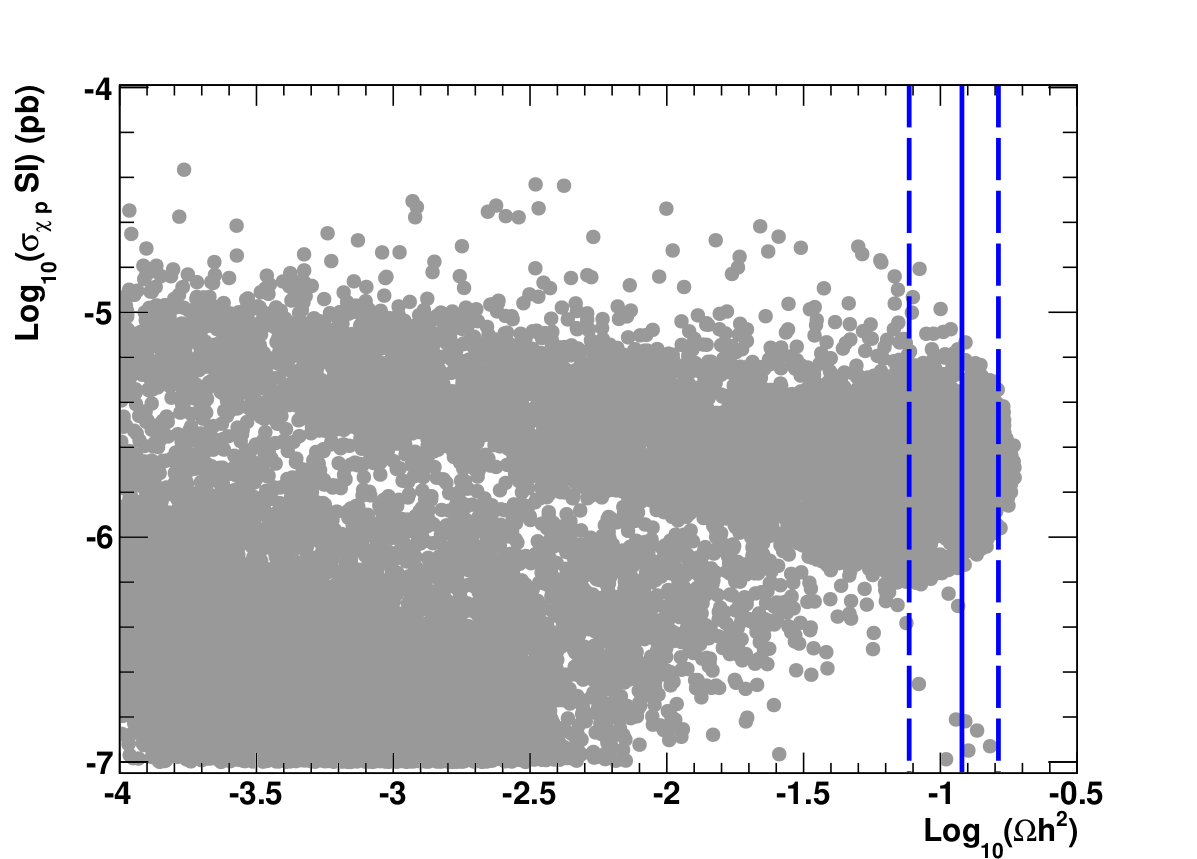}
\vspace*{-0.40cm}
\caption{Scattering cross section as a function of the neutralino relic density for pMSSM points passing all other selection criteria used 
in this study. The vertical lines show the PLANCK relic DM density value and the range of the tight constraint applied here.}
\label{fig:Oh2}
\end{figure}


The correlation of the neutralino relic density and the spin-independent $\tilde{\chi}-p$ scattering cross section is important, as highlighted in Fig.~\ref{fig:Oh2}. 
In general, points with large scattering cross section correspond to small values of neutralino relic density, due to the fact that in this 
region, the splitting between the neutralino and the sbottom is small, resulting in an increased annihilation cross section. 
However, after applying all other constraints the points selected by the relic density have relatively high scattering cross section.
In the following we consider both loose, $10^{-4} < \Omega_{\chi} h^2 < 0.163$, and tight, 
$0.076 < \Omega_{\chi} h^2 < 0.163$, neutralino relic density constraints.

\subsection{Indirect Detection and Other Constraints}

Indirect detection experiments provide us with constraints on DM by
analysing the cosmic ray fluxes. In particular, PAMELA
\cite{Adriani:2008zr}, FERMI \cite{FermiLAT:2011ab}, HESS
\cite{Aharonian:2009ah} and AMS-02 \cite{Aguilar:2013qda} have detected
excesses in the electron-positron spectra, while PAMELA has a precise
measurement of the antiproton flux~\cite{Adriani:2010rc} which does not
reveal any excess compared to predictions. FERMI-LAT has also released
strong bounds from $\gamma$-ray searches~\cite{Ackermann:2011wa}. While
the $e^\pm$ excesses could be interpreted in terms of DM, the general
accepted explanation is in terms of astrophysical phenomena
\cite{Cirelli:2012tf}. Here, we adopt the upper limit on the
$\tilde{\chi}$ annihilation cross sections derived in
\cite{DeSimone:2013fia,Cirelli:2013hv} as a constraint. The strongest
limits on annihilation cross sections come from the FERMI-LAT
$\gamma$-ray searches.
In our specific model with light neutralinos and light sbottoms, the
main annihilation channel is either $\tilde\chi^0_1 \tilde\chi^0_1 \to
b\bar{b}$, mediated by a $Z$ or Higgs boson in $s$-channel, or by a
sbottom in $t$-channel, or the 3-body decay $\tilde\chi^0_1
\tilde\chi^0_1 \to b\bar{b} g$. However, due to the suppressed couplings
of the lightest neutralino to the $Z$ and $h$, the $s$-channel is
suppressed and the annihilation cross section is expected to be small.
In the WIMP mass region of interest to our analysis, the strongest bound
on the total annihilation cross sections times velocity is $\sim
10^{-26}$ cm$^3$/s, obtained by FERMI-LAT from gamma-flux measurements,
while the upper bound on the $\tilde\chi^0_1 \tilde\chi^0_1 \to b \bar{b} g$ 
cross-section is $\sim 2 \times 10^{-27}$cm$^3$/s, obtained from
PAMELA antiproton flux measurements \cite{Asano:2011ik}.

\begin{figure}[h!]
\includegraphics[width=0.65\columnwidth]{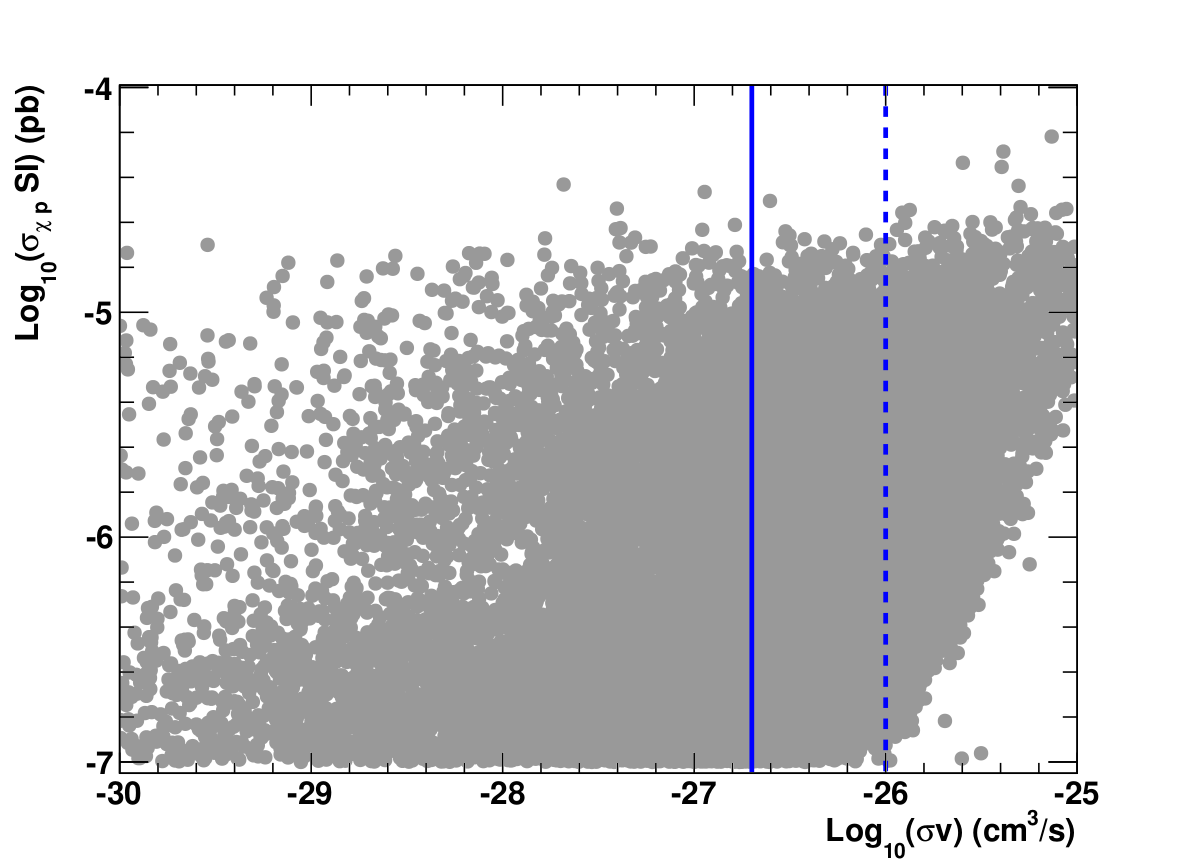}
\vspace*{-0.40cm}
\caption{Scattering cross section as a function of the total DM annihilation cross section for 
selected pMSSM points. 
The vertical dashed and solid lines show the Fermi-LAT $\gamma$-ray strongest upper limit on the 
$\tilde{\chi} \tilde{\chi} \rightarrow b \bar b$ and the PAMELA $\bar p$ strongest upper limit on 
$\tilde{\chi} \tilde{\chi} \rightarrow b \bar b g$ annihilation cross sections, respectively.}
\label{fig:ID}
\end{figure}

We calculate annihilation cross sections with a modified version of {\tt
micrOMEGAs}, which includes the calculation of the $\tilde\chi^0_1 \tilde\chi^0_1 \to b\bar{b} g$ cross-section.
Fig. \ref{fig:ID} shows the correlation
between the total annihilation cross section and the scattering cross
section. Again, we see that more points with large scattering cross
sections are present in the region with large annihilation cross
sections. Also, most of the points are below the upper limit on the
annihilation cross section.

\indent In addition to the constraints described in the previous subsections, we checked that the running of $\alpha_s$ is not affected by the presence of light sbottoms since the gluino is heavy in our scenario~\cite{Chiang:2002wi} as also mentioned in~\cite{Gondolo:2013wwa}. Furthermore, we consider constraints from flavour physics and muon anomalous magnetic moment as detailed in Table~\ref{tab:constraints}, which summarises 
all constraints applied in our analysis.
\begin{table*}[t!]
\begin{center}
\begin{tabular}{|c|c|}
\hline
Type & Constraint\\
\hline\hline
Higgs mass constraint & $M_h \in [121,129]$ GeV\\
Higgs signal strengths & Table \ref{tab:input}\\
\hline
\multirow{3}{*}{$Z$ decay widths} & $\Gamma(Z\to \tilde{\chi}^0_1 \tilde{\chi}^0_1) < 3$ MeV\\
 & $\Gamma(Z\to \tilde{\chi}^0_1 \tilde{\chi}^0_1) +\Gamma(Z\to \tilde{b}_1 \tilde{b}_1)  < 5$ MeV\\
 & $0.21497 < R_b < 0.21761$\\
\hline
\multirow{2}{*}{LEP and Tevatron SUSY searches} & as given in \cite{Arbey:2012na}\\
 & + specific analysis of the $\tilde{\chi}^+ \tilde{\chi}^- / \tilde{\chi}_2^0 \tilde{\chi}_1^0$ channels \\
\hline
Oblique parameters $S$, $T$, $U$ & LEP, see Fig.~\ref{fig:stu}\\
\hline
Vacuum stability & stable or long-lived scalar potential minimum\\
\hline
\multirow{6}{*}{Flavour physics} & $2.63\times10^{-4}<\mbox{BR}(B\to X_s\gamma) < 4.23\times10^{-4}$ \cite{Amhis:2012bh}\\
& $1.28\times10^{-9}<\mbox{BR}(B_s\to \mu^+\mu^-)_{\rm untag} < 4.52\times10^{-9}$ \cite{Aaij:2013aka,Chatrchyan:2013bka,bsmm-comb}\\
& $0.40\times10^{-4}<\mbox{BR}(B_u\to \tau\nu) < 1.88\times10^{-4}$ \cite{Adachi:2012mm,Lees:2012ju}\\
& $4.7\times10^{-2}<\mbox{BR}(D_s\to \tau\nu) < 6.1\times10^{-2}$ \cite{Amhis:2012bh,Akeroyd:2009tn}\\
& $2.9\times10^{-3}<\mbox{BR}(B\to D^0\tau\nu) < 14.2\times10^{-3}$ \cite{Aubert:2007dsa}\\
& $0.985 < R_{\mu23} < 1.013$ \cite{Antonelli:2008jg}\\
\hline
Muon anomalous magnetic moment & $-2.4\times10^{-9}< \delta a_\mu < 4.5\times10^{-9}$ \cite{Arbey:2011un} \\
\hline
Loose relic density & $10^{-4} < \Omega_{\chi} h^2 < 0.163$\\
Tight relic density & $0.076 < \Omega_{\chi} h^2 < 0.163$\\
\hline
\multirow{2}{*}{Dark matter annihilation cross-section} & $\sigma v_{\rm tot} < 10^{-26}$ cm$^3$/s with $M_{\tilde{\chi}_1^0} < 50$ GeV \cite{Ackermann:2011wa}\\
& $\sigma v_{bbg} < 2 \times 10^{-27}$ cm$^3$/s with $M_{\tilde{\chi}_1^0} < 50$ GeV \cite{Adriani:2010rc,Asano:2011ik}\\
\hline
\multirow{2}{*}{Dark matter direct detection} & $10^{-7} <\sigma^{\rm SI}_{p-\chi}< 10^{-2}$ pb with $M_{\tilde{\chi}_1^0} < 50$ GeV\\
& (close to the CDMS contour and XENON limit)\\
\hline
\multirow{3}{*}{LHC searches} & Higgs searches\\
& SUSY searches\\
& $pp \rightarrow \chi \chi +$ jet, $\gamma$ and $Z/W$ searches\\
\hline
\end{tabular}
\end{center}
\caption{Summary of the constraints.}
\label{tab:constraints} 
\end{table*}

\section{Dark Matter and LHC Constraints}

\subsection{Light Dark Matter and the Higgs}

The discovery of a light Higgs-like particle and the first determinations of its mass and couplings have important consequences for light DM scenarios. In a generic DM model, the coupling of the Higgs to the WIMP particles is responsible for the correlation between DM and Higgs sectors.
First, the Higgs boson contributes to both the WIMP scattering cross section and $\tilde{\chi} \tilde{\chi}$ annihilation processes.
Then, if the lightest neutralino and possible other SUSY particles exist at masses smaller than $M_h/2$, the lightest Higgs boson decays into pairs of these particles, in particular $h \rightarrow \tilde{\chi} \tilde{\chi}$.

If the $h\tilde{\chi}\tilde{\chi}$ coupling is large, and the splitting of the neutralino with the other SUSY particles is large, which corresponds to a gaugino mixed state, the rate for the invisible decay $h \to \tilde{\chi}^0_1 \tilde{\chi}^0_1$, if kinematically allowed, can be large. In this case, the Higgs impact on DM direct detection searches is important. The lightest Higgs boson can mediate the scattering with nucleons, and modify the scattering cross-section which is normally mediated by a $Z$ boson. 
The enhanced coupling of neutralinos to the Higgs opens an annihilation channel, which increases the effective cross-section and decreases the neutralino relic density. 

On the other hand, if the coupling of the neutralino to the Higgs is large, but other light SUSY particles exist, these may blur the correlations between the Higgs and DM sectors. Scenarios where the decay $h \to \tilde{\chi}^0_1 \tilde{\chi}^0_1$ is open, the Higgs decays to other SUSY particles or the decays to SM particles are modified, can be strongly constrained by the LHC Higgs results. DM searches may be more widely affected. For DM direct searches, the presence of a light SUSY scalar provides an additional $t$-channel mediation. For the neutralino relic density, DM co-annihilations can increase the effective annihilation cross section and decrease the final neutralino density. Finally, in DM indirect searches, the annihilation channels can be mediated by the additional light SUSY particles in $t$-channels. 

\begin{figure}[t!]
\begin{tabular}{cc}
\includegraphics[width=0.5\columnwidth]{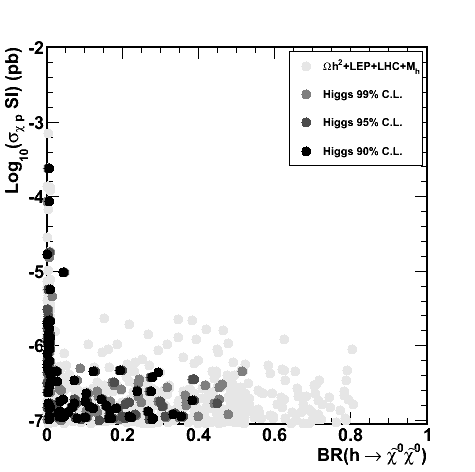} &
\includegraphics[width=0.52\columnwidth]{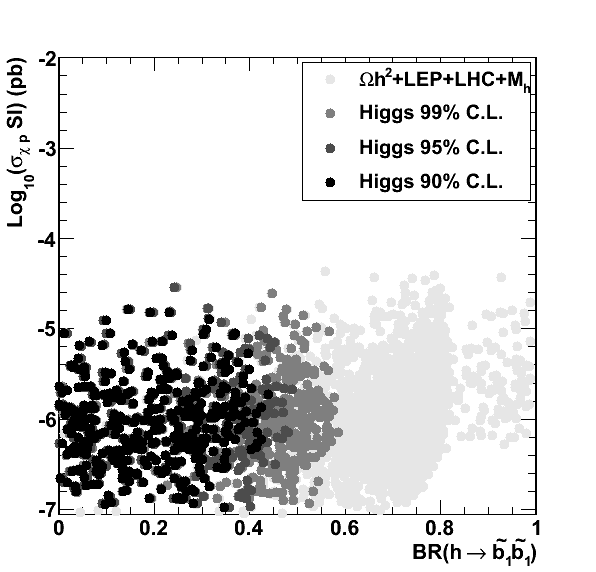} \\
\end{tabular}
\vspace*{-0.40cm}
\caption{Scattering cross section as a function of the lightest Higgs decay branching fractions for $h \rightarrow \tilde{\chi}^0_1 \tilde{\chi}^0_1$ (left) and $h \rightarrow \tilde{b}_1 \tilde{b}_1$ (right). The points 
corresponding to the light neutralino, almost degenerate sbottom, cluster at the smallest values of the invisible branching fraction. The colour scale denotes the compatibility with the LHC Higgs data.}
\label{fig:BRh}
\end{figure}
\begin{figure}[t!]
\begin{tabular}{cc}
\includegraphics[width=0.5\columnwidth]{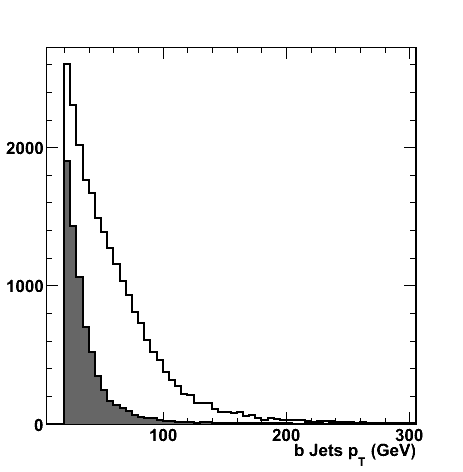} &
\includegraphics[width=0.52\columnwidth]{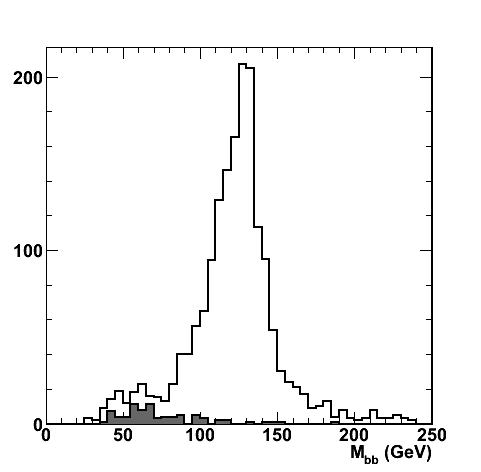} \\
\end{tabular}
\vspace*{-0.40cm}
\caption{Observables in Higgs decay searches: (left) $b$-jet transverse momentum and (right) di-jet invariant mass. The shaded histogram represents  $h \rightarrow \tilde{b}_1 \tilde{b}_1$ and the open histogram $h \rightarrow b \bar b$.}
\label{fig:hb1b1}
\end{figure}

Instead, if the lightest neutralino is a nearly pure gaugino state, the Higgs does not couple to the neutralino, and the decay $h \to \tilde{\chi}^0_1 \tilde{\chi}^0_1$ is either suppressed or completely forbidden. In this case, the presence of the Higgs has only an effect if resonances in neutralino annihilations mediated by Higgs appear, but the coupling suppression reduces the correlation between the Higgs and DM sectors. This is the case for our selected pMSSM points, where 
the decay to two neutralinos is strongly suppressed, corresponding to a pure-bino $\tilde{\chi}^0_1$ (see Figure~\ref{fig:BRh}).
The low mass of the lightest scalar bottom makes possible to get a sizeable Higgs decay rate into $\tilde{b}_1 \tilde{b}_1$ pairs, which would represent an 
important signal for the LHC data (see Figure~\ref{fig:BRh}), as already mentioned in~\cite{Arbey:2012na}. Now, we study this process in detail by contrasting it with 
the $h \rightarrow b \bar b$ decay channel. In fact, the decay $h \rightarrow \tilde{b}_1 \tilde{b}_1$ leads to a final state similar to that of 
$h \rightarrow b \bar b$, but with MET from the two escaping neutralinos. In order to evaluate the contribution of $h \rightarrow \tilde{b}_1 \tilde{b}_1$ into the search region for the $b \bar b$ channel, we study the reconstruction of both decays in events generated with {\tt Pythia 8.150}~\cite{pythia8} using the {\tt Delphes 3.0} fast simulation. Jets are reconstructed using the anti-kt algorithm~\cite{Cacciari:2008gp}, 
implemented in the {\tt FastJet} package~\cite{fastjet}, with a cut of 0.4. The invariant mass of pairs of jets associated to a $b$ quark and having $p_{T}>25$ GeV, $M_{bb}$ is computed. We obtain a di-jet mass resolution $\delta M/M \sim 0.13$ for the $b \bar b$ channel, which agrees well with the performance obtained on full simulation for the LHC $H_{SM} \rightarrow b \bar b$ searches. $h \rightarrow \tilde{b}_1 \tilde{b}_1$ decays have a $M_{bb}$ distribution peaked around 50~GeV due to the loss of the two neutralinos and a tail extending up to the Higgs mass region, which accounts for $\sim$25\% of the reconstructed events. In Fig.~\ref{fig:hb1b1} the $b$-jet transverse momentum and di-jet invariant mass are shown for $h \rightarrow \tilde{b}_1 \tilde{b}_1$ and $h \rightarrow b \bar b$. Due to the softer $b$ jets, the acceptance of $h \rightarrow \tilde{b}_1 \tilde{b}_1$ in the $Wh$, $h \rightarrow b \bar b$ analysis is very limited and we estimate from the fast simulation study that only 10\% of such events would be selected in the $bb$ signal region. Therefore, a sizeable $\tilde{b}_1 \tilde{b}_1$ rate would induce a reduction of the signal strengths in the other modes. This feature may provide
a good opportunity for a test at the LHC, once the $h \rightarrow  \bar{b}b$ decay will have been established and its signal strength measured with sufficient precision.

Finally, we have analysed the pMSSM points corresponding to the light, almost degenerate neutralino-sbottom scenario for their compatibility with the present LHC Higgs data. 
For each point we compute the $\chi^2$ probability comparing the Higgs mass and signal strengths for the pMSSM point to the LHC measurements given in Table~\ref{tab:input}. 
\begin{table}[!h]
\begin{center}
\begin{tabular}{|c|c|c|}
\hline
Parameter & Value & Experiment \\ \hline \hline
$M_h$ (GeV)   & 125.7$\pm$0.4 & ATLAS\cite{ATLAS-CONF-2013-014}+CMS\cite{CMS-13-002} \\ 
$\mu_{\gamma \gamma}$ & 1.20$\pm$0.30 & ATLAS\cite{ATLAS-CONF-2013-012}+CMS\cite{CMS-13-001} \\
$\mu_{Z Z}$ & 1.10$\pm$0.22 & ATLAS\cite{ATLAS-CONF-2013-013}+CMS\cite{CMS-13-002} \\
$\mu_{W W}$ & 0.77$\pm$0.21 & ATLAS\cite{ATLAS-CONF-2013-030}+CMS\cite{CMS-13-003} \\ 
\hline
\end{tabular}
\end{center}
\caption{Input average values of the $h$ mass and signal strengths used for this study with their statistical accuracies.}
\label{tab:input} 
\end{table}
In the following, selected pMSSM points are classified according to their compatibility with the LHC Higgs data based on the observed $\chi^2$ probability.  

\subsection{SUSY Searches}

\begin{figure}[t!]
\begin{tabular}{cc}
\includegraphics[width=0.5\columnwidth]{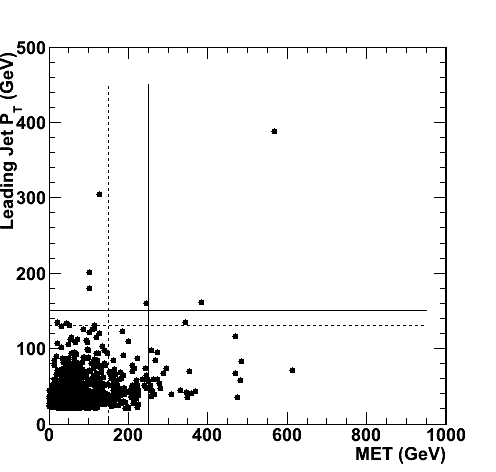} &
\includegraphics[width=0.5\columnwidth]{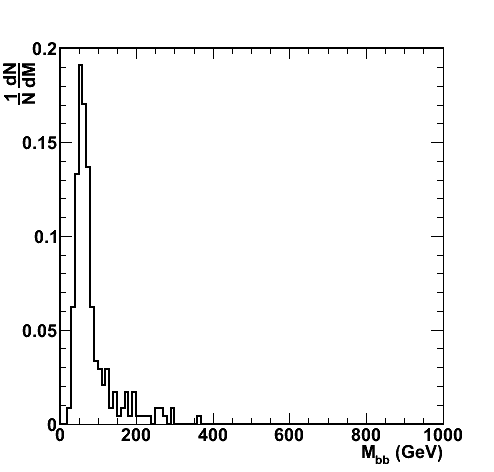} \\
\end{tabular}
\vspace*{-0.40cm}
\caption{Observables in jets + MET SUSY searches: (left) largest transverse momentum of $b$-jet in the event vs.\ 
MET and (right) di-jet invariant mass in $pp \rightarrow \tilde{b}_1 \tilde{b}_1$. The lines indicate the cuts adopted for the signal selection of the ATLAS analysis.}

\label{fig:sbottom}
\end{figure}

The LHC has extensively searched for SUSY particle production in hadronic and leptonic final states with significant MET. Most relevant here are the direct searches 
for scalar bottom pair and weakly interacting sparticle production. 
In general, the $pp \rightarrow \tilde{b}_1 \tilde{b}_1$ escapes detection in the LHC SUSY analysis with jets + MET due to the small jet $p_T$ and low MET, despite its 
large cross section. In order to study in details the possible sensitivity to direct scalar bottom production at the LHC, samples of these events have been generated 
with {\tt Pythia 8.150} at 8~TeV with the CTEQ6L1 parton distribution functions (PDFs)~\cite{Pumplin:2002vw} and analysed using {\tt Delphes 3}. The same signal selections 
as in the preliminary ATLAS~\cite{ATLAS-CONF-2013-053} and CMS~\cite{CMS-SUS-12-028} analyses have been adopted. In particular, the ATLAS analyses use a specific event 
selection optimised for small mass splitting, which was not available when we performed our earlier study of~\cite{Arbey:2012na}. 
Figure~\ref{fig:sbottom} shows the $b$-jet transverse energy, the MET and di-jet invariant mass for the $pp \rightarrow \tilde{b}_1 \tilde{b}_1$ events, compared to the 
cuts applied in the ATLAS analysis. An efficiency of only $\sim$2$\times$10$^{-5}$ for $\tilde{b}_1 \tilde{b}_1$ is obtained, due to the relatively high cuts on MET and 
$H_T$ applied in the ATLAS and CMS analyses, respectively.

We also ensure that the selected pMSSM points of our scenario are not excluded by the searches for other SUSY particles. 
Both ATLAS and CMS have searched for chargino and neutralino production in multi-lepton final states. These searches are sensitive to 
$pp \rightarrow \tilde{\chi}^{\pm}_1 \tilde{\chi}^0_2$ with the subsequent decays $\tilde{\chi}^{\pm}_1 \rightarrow W^{\pm} \tilde{\chi}^{0}_1$ 
and $\tilde{\chi}^0_2 \rightarrow \tilde{\ell} \ell$, $Z \tilde{\chi}^{0}_1$. Since their sensitivity depends on particle masses, decay branching fraction 
and mass patterns, we test the observability of these processes in our points by generating samples of signal events with {\tt Pythia 8.150}
at 8~TeV for each of the selected pMSSM points and use {\tt Delphes 3} for reconstructing the physics objects. We then apply the selection criteria of the ATLAS preliminary 
analyses of~\cite{ATLAS-CONF-2013-035,ATLAS-CONF-2013-049} to the reconstructed events and compare the number of selected signal events to that of background 
events obtained in the ATLAS data analysis. The 95\% confidence level exclusion of each SUSY point in presence of background only is determined using the CLs 
method~\cite{Read:2002hq}. 
However, in the light neutralino, almost degenerate $\tilde{b}_1$ scenario the $\tilde{\chi}^0_2$ decays preferentially to $\tilde{b}_1 b$ giving 
a $\ell$ + $b b$ + MET topology, which should be investigated at the LHC at 13-14~TeV.  

For the scalar top searches, we compute the product of $pp \rightarrow \tilde{t}_1 \tilde{t}_1$ production cross sections and 
$\tilde{t}_1 \rightarrow t \tilde{\chi}^{0}_1$, $\tilde{t}_1 \rightarrow b \tilde{\chi}^{\pm}_1$ + 
$\tilde{\chi}^{\pm}_1 \rightarrow W^{\pm} \tilde{\chi}^{0}_1$ decay branching fractions to those excluded in the preliminary ATLAS analyses 
of~\cite{ATLAS-CONF-2013-037,ATLAS-CONF-2013-048}.

\subsection{Mono jet, $W$, $Z$ Searches}

Hadron colliders are sensitive to DM coupling to quarks and it is in general 
possible to establish a relation between the $\tilde{\chi}$ -- nucleon
scattering cross section and the rate of production of events with a
jet, photon or gauge boson and transverse momentum imbalancement~\cite{Bai:2010hh}. 
The CDF and D0 experiments at Tevatron have already searched for mono-jet 
events~\cite{Abazov:2003gp,Aaltonen:2012jb}.
The ATLAS and CMS experiments have searched for the processes $pp \rightarrow \tilde{\chi} \tilde{\chi} X$, 
with $X$ being a hadronic jet~\cite{ATLAS-COM-CONF-2012-190,CMS-EXO-12-048} and a single photon~\cite{Chatrchyan:2012tea,Aad:2012fw}.
More recently, the ATLAS experiment has performed a similar analysis for
hadronically-decaying $W$ and $Z$ bosons~\cite{ATLAS-CONF-2013-073}.
Currently the best sensitivity has been reported by CMS from the search
of mono-jet events on 20~fb$^{-1}$ of 8~TeV data~\cite{CMS-EXO-12-048}.
The results of this analysis are interpreted as an upper limit on the DM -- nucleon spin-independent 
scattering cross section of 1.24 $\times$ 10$^{-39}$~cm$^{-2}$ for a WIMP mass 
of 10~GeV and axial vector operator. This limit is still several orders of
magnitude away from the region highlighted by the recent CDMS result and
characteristic of the points selected from our pMSSM scans, which are in the range 
10$^{-43}$-10$^{-42}$~cm$^{-2}$.

However, special care should be taken in interpreting these limits in the context of SUSY, since they are derived under 
the assumption that only one operator contributes in the amplitudes and only one dark matter particle and one mediator 
are involved. In addition, it is assumed that the mediator does not couple to gauge bosons and the coupling to the Higgs 
is also negligible~\cite{Goodman:2010ku}. These assumptions do not hold for SUSY, in general, and for our scenario, 
in particular.

\begin{figure}[t!]
\begin{tabular}{cc}
\includegraphics[width=0.5\columnwidth]{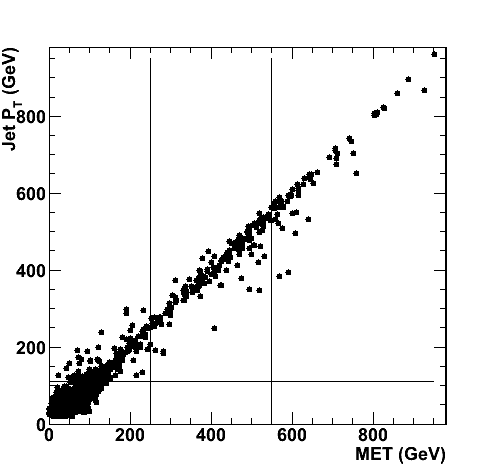} &
\includegraphics[width=0.5\columnwidth]{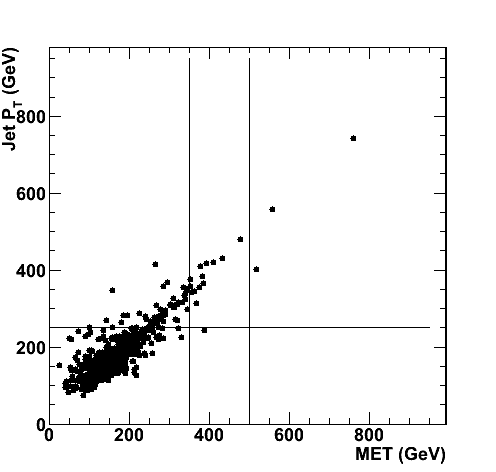} \\
\end{tabular}
\vspace*{-0.40cm}
\caption{Largest transverse momentum of hadronic jet in the event vs.\ MET for 
$pp \rightarrow \tilde{\chi}\tilde{\chi} {\mathrm{Jet}}$ (left) and $pp \rightarrow \tilde{\chi}\tilde{\chi} W/Z$ (right) searches. 
The lines indicate the cuts adopted for defining the signal regions in the CMS and ATLAS analyses, respectively.}
\label{fig:mono}
\end{figure}

In our earlier study we imposed the value of the upper limit derived by CMS on 5~fb$^{-1}$ at 7~TeV for the DM -- nucleon spin-independent 
scattering cross section~\cite{Chatrchyan:2012me} as a constraint on the pMSSM points.
In order to test more precisely the sensitivity  of these searches to our scenario, here we explicitly study the acceptance 
of $pp \rightarrow \tilde{\chi}\tilde{\chi} +$ jet, $pp \rightarrow \tilde{b}_1 \tilde{b}_1 +$ jet events by the CMS 
mono-jet~\cite{CMS-EXO-12-048} and $pp \rightarrow \tilde{\chi}\tilde{\chi} + W$ or $Z$ events by the ATLAS single $W$ and $Z$ 
boson~\cite{ATLAS-CONF-2013-073} analyses performed on the 8~TeV LHC data. 
We compute the cross section for $pp \to \tilde{b}_1 (\to b \tilde{\chi}^0_1) \bar{\tilde{b}}_1 (\to \bar{b} \tilde{\chi}^0_1) +
{\rm jet}$ using $\tt MadGraph~5$. We paid attention to include the full
$2 \to 5$ matrix elements, as the narrow width approximation is known to
break down in the regions where the daughter particles are close in mass
to the parent particle~\cite{Berdine:2007uv}, which is typically the
case of a scalar bottom of $\sim 15$ GeV decaying to a neutralino and bottom
of $\sim$10 and $\sim$5 GeV respectively. Using the narrow width approximation would incorrectly
increase the resulting cross sections by large factors and lead to
erroneous conclusions in our scenario.
Signal samples in the jet, $Z$ and $W$ + MET are generated
using {\tt MadGraph 5} for a set of selected pMSSM points, through
the corresponding SLHA files. Events are hadronised using {\tt Pythia 8.150}
with the CTEQ6L1 PDF set and then processed through {\tt Delphes 3} to
obtain the reconstructed physics objects for the subsequent analysis. 
The production cross section is $\sim$0.5 - 0.2~fb for the case of $pp \rightarrow \tilde{\chi}\tilde{\chi} +$ jet 
and $\sim$9 - 1~fb for $W+Z$ events, depending on the neutralino mass. 
Jets are reconstructed using the anti-kt algorithm 
with a distance parameter of 0.5 for the mono-jet search and the
Cambridge-Aachen algorithm~\cite{Dokshitzer:1997in} with a radius
parameter of 1.2 for the $W/Z$ search, to reproduce the procedure of the
original analyses. Figure~\ref{fig:mono} shows the correlation of the
transverse momentum of the leading $p_T$ jet with the event MET, where
only jets with invariant mass in the range 50 to 120~GeV are accepted
for the $W+Z$ analysis.

The fraction of  $pp \rightarrow \tilde{\chi}\tilde{\chi} +$ jet and $W$ or $Z$ events fulfilling the loose (tight) signal 
cuts adopted in the original analyses is 4.5$\times$10$^{-2}$ (1.2$\times$10$^{-2}$) for the mono-jet search and 
9.5$\times$10$^{-4}$ (1.5$\times$10$^{-4}$) for the $W+Z$ search. Comparing to the excluded product of production cross section 
and reconstruction efficiency for the CMS and ATLAS analyses, our pMSSM points are well below the bounds established 
with 20~fb$^{-1}$ of 8~TeV data. 

In the case of the $pp \rightarrow \tilde{b}_1\tilde{b}_1 +$ jet process, the cross sections for the selected pMSSM points 
corresponding to our light $\tilde{b}_1$ scenario are significantly larger, but the efficiency of the CMS analysis cuts is 
also smaller. We compute the cross section for the $pp \rightarrow b \tilde{\chi}^0_1 \bar{b} \tilde{\chi}^0_1$ + jet using 
{\tt MadGraph 5} and obtain values in the range 1000 to 200~pb for 11 $< M_{\tilde{b}_1} <$ 35~GeV. The 
efficiency for obtaining events with a hard jet with $p_T >$ 80~GeV and MET above 300 (350) GeV is in the range 
1.4$\times 10^{-5}$ - 9$\times 10^{-4}$ (4.5$\times 10^{-6}$ -  3$\times 10^{-4}$), with the efficiency increasing with the 
$\tilde{b}_1$ mass and the $\Delta M$ values. These small values of efficiency, depending on the extreme tail of the MET 
distribution, need to be confirmed by a detailed detector simulation. However, taken at face value they imply an exclusion 
of only the pMSSM points with $M_{\tilde{\chi}^0_1} >$ 24~GeV, thus allowing the bulk of the region consistent with CDMS and 
other data.   

\section{Discussion}

\begin{figure}[b!]
\vspace*{-0.50cm}
\includegraphics[width=0.70\columnwidth]{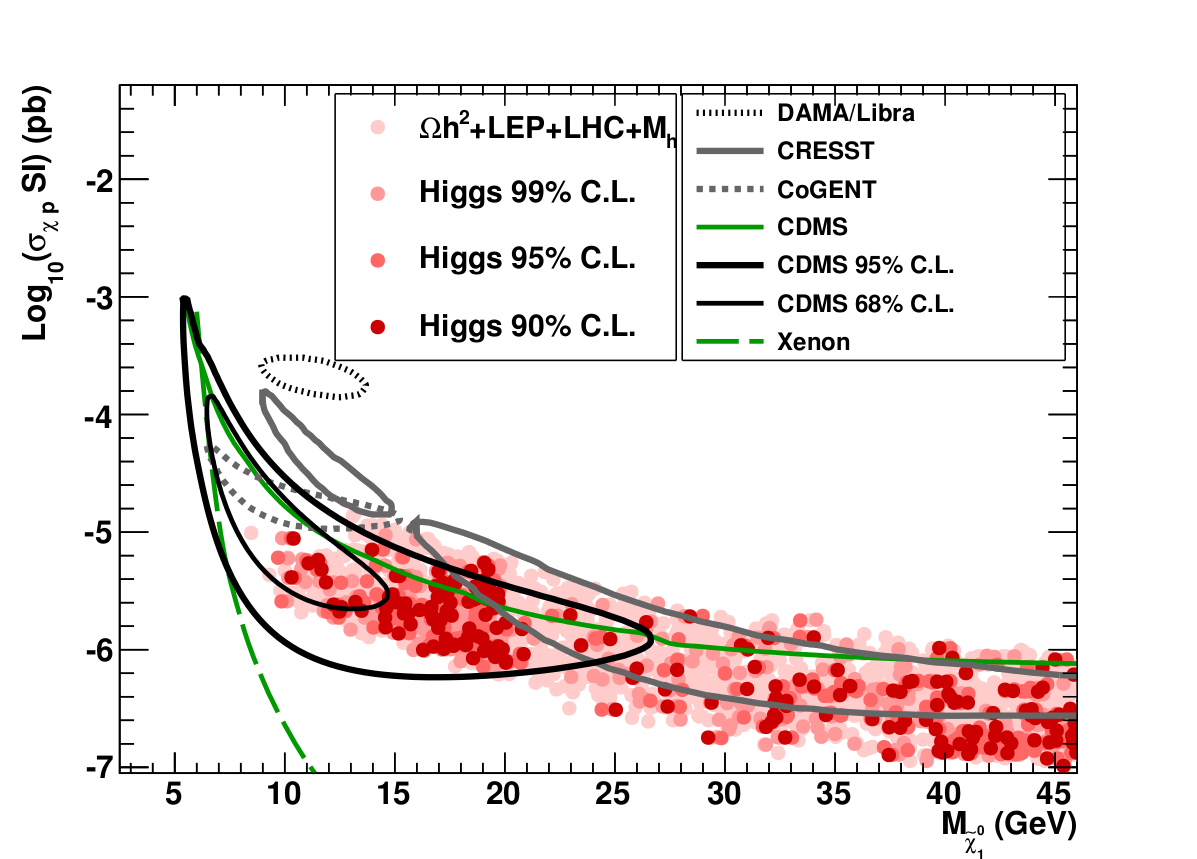} \\
\vspace*{-0.05cm}
\includegraphics[width=0.70\columnwidth]{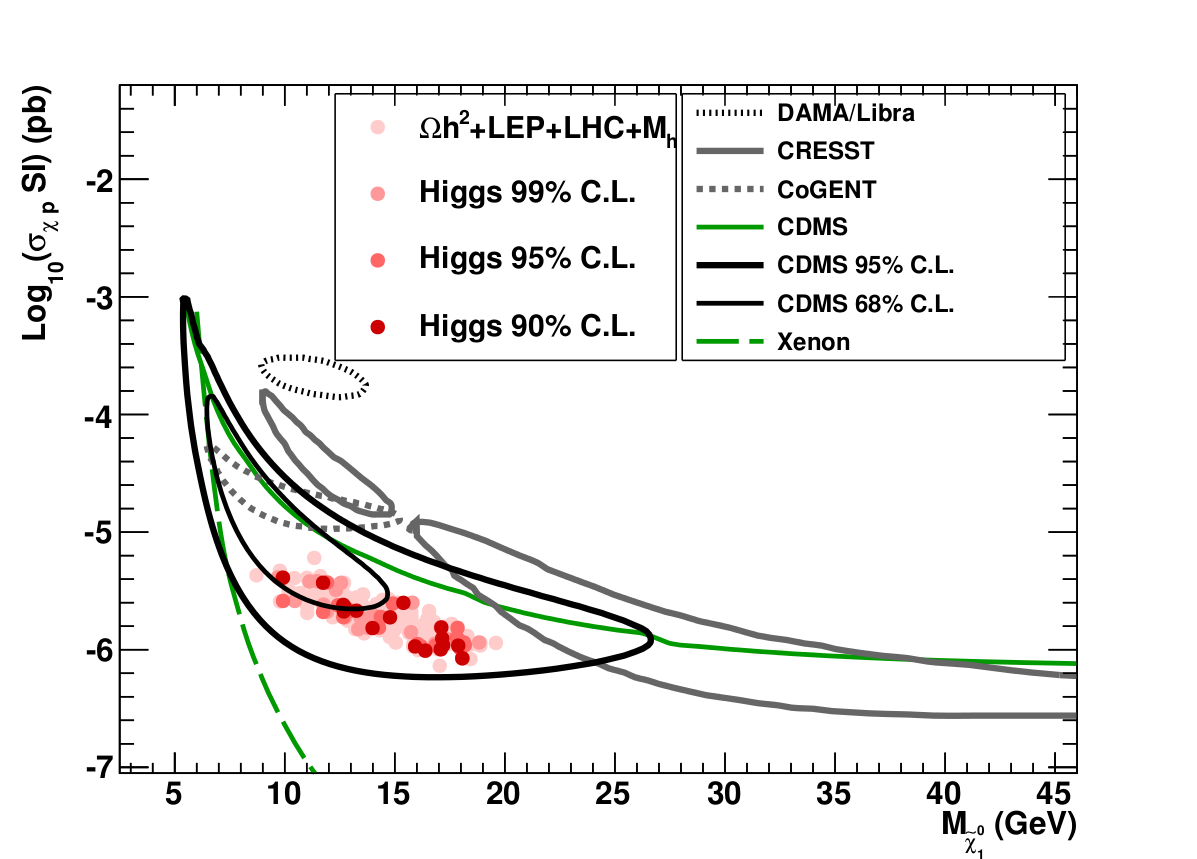} 
\vspace*{-0.40cm}
\caption{Scattering cross section as a function of the lightest neutralino mass. The points represent pMSSM solutions from our scans and the colour scale gives their compatibility with the Higgs signal strengths obtained by ATLAS and CMS. The lines indicates the regions corresponding to results of direct DM searches. In the upper plot, 
the loose neutralino relic density constraint, $10^{-4} < \Omega_{\chi} h^2 < 0.163$, is applied, while the lower plot uses the tight constraint, 
$0.076 < \Omega_{\chi} h^2 < 0.163$. The region corresponding to $M_{\tilde{\chi}^0_1} >$ 24~GeV is excluded by the CMS mono-jet analysis, according 
to our estimate as discussed in the text.}
\label{fig:MN1DDOh2}
\end{figure}

After applying the constraints discussed above, the selected points from our scan provide MSSM solutions with neutralino LSP compatible 
with the light WIMP scenario suggested by the tantalising recent CDMS result and by other DM direct detection experiments. Fig.~\ref{fig:MN1DDOh2} shows 
the $\tilde{\chi}-p$ scattering cross section as a function of the $\tilde{\chi}$ mass for our selected pMSSM points compared to the 
results of the direct detection experiments. 
These points, corresponding to the light neutralino, almost degenerate $\tilde{b}_1$ scenario are consistent with the recent PLANCK result 
on the relic DM density, interpreted either as an upper limit on the neutralino relic density or as a tight constraint (as defined in 
Table~\ref{tab:constraints}). Of these points, 6$\times$10$^{-4}$ are compatible at 90\% C.L. with the LHC Higgs data of Table~\ref{tab:input}. 
The preliminary results of the CMS mono-jet search at 8~TeV leave the bulk of the region consistent with CDMS and other data unaffected, 
according to our analysis as discussed above.   
 
It is remarkable that this appears to be the only scenario in the MSSM with neutralino LSP providing us with 
a light neutralino LSP, with mass below 20~GeV, and large WIMP-nucleon scattering cross section. The large squark 
mixing angle $\theta_b$, close to $\pi/2$, makes the $\tilde{b}_1$, mostly $\tilde{b}_R$, almost degenerate 
with the $\tilde{\chi}^0_1$ LSP and reduces its coupling to the $Z$, thus ensuring compliance with the $Z$ lineshape 
and the $e^+e^-$ searches. On the basis of the result of the pMSSM scans and fast simulation studies we have reported here 
a light neutralino with mass below $\sim 20-30$ GeV is not yet experimentally excluded, as suggested by~\cite{Han:2013gba}.
This discrepancy in the results is due to the different coverage of the pMSSM parameter space of the two studies, in particular 
the fact that only scenarios with scalar quark masses larger than 100~GeV are retained in the analysis of ~\cite{Han:2013gba}.

\begin{figure}[bh!]
\begin{center}
\vspace*{-0.25cm}
\includegraphics[width=0.75\columnwidth]{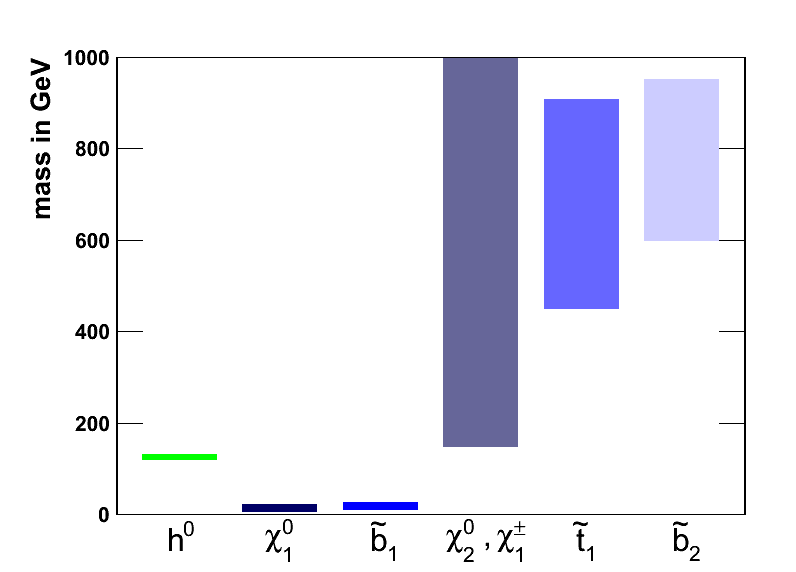}
\vspace*{-0.85cm}
\end{center}
\caption{Range of the masses of the relevant SUSY particles in the pMSSM scenario with a light neutralino and 
nearly degenerate $\tilde{b}_1$. \label{spectrum}}
\end{figure}

The pMSSM parameters most important in defining the spectrum for this scenario are $M_1$, $M_{Q3_L}$, $M_{b_R}$, $A_b$ and $A_t$ while 
$\mu$ needs to be large. 
Fig.~\ref{spectrum} summarises the ranges of the relevant particles. Masses of the other SUSY particles can be pushed to large values 
without affecting the viability of the MSSM solution for the constraints discussed in this paper.  There are some important common 
features. First, the LSP is a bino-like neutralino of mass $\sim 10 - 20$ GeV, and the NLSP is a $\tilde{b}_1$ mainly right-handed of 
mass $\sim 15 - 25$ GeV. The $\tilde{\chi}^0_2$ and $\tilde{\chi}^\pm_1$ have masses $\ge$150~GeV and are wino-like or mixed states 
depending on their masses.  The mass of the lightest Higgs is $\sim 126$ GeV to agree with the LHC results and the $\tilde{\chi}^0_1$ 
has very strongly suppressed couplings to the Higgs and $Z$ bosons, while the $\tilde{b}_1$ couples only weakly to them. 
This ensures that the scenario is not excluded by electroweak data and it does not produce a significant invisible decay width of the 
$Z$ and $h$ bosons. The branching ratio for the decay 
$h \to \tilde{b}_1 \tilde{b}_1$ ranges from values as low as a few \% up to 50\%, or more. The $\tilde{t}_1$ and $\tilde{b}_2$ are 
constrained by the common value of third generation scalar quark masses in the pMSSM and the $\tilde{t}_1$ contribution to the $h$ mass. 
Still, the $\tilde{t}_1$ mass can be safely chosen to be above 600~GeV, beyond the current reach of the LHC searches. We observe in 
passing that points in the region where the scalar top mass exceeds 700~GeV generally fail the vacuum stability test, however this feature 
may be specific of the points obtained in our scans and heavier stop masses may be allowed for other choices of the pMSSM parameters.

\section{Conclusions}

The MSSM offers solutions compatible with a light WIMP and large scattering cross section, as suggested by CDMS and other data, if the 
reported events are due to a DM signal. The light, almost degenerate sbottom scenario remains a viable solution when the constraints from  
$e^+e^-$ experiments and from the latest LHC results are applied. In this scenario, a sizeable $h \rightarrow \tilde{b}_1 \tilde{b}_1$ rate 
may provide a good opportunity for a test at the LHC once the $h \rightarrow  b \bar{b}$ decay will have been established and its signal 
strength measured with sufficient precision. Interesting opportunities for dedicated searches of light sbottoms arise at LHC, in the 
$b$-jets + MET, if the kinematical cuts can be lowered, and mono-jet channels as well as at a future $e^+e^-$ collider and should be pursued, 
if the first, tantalising indications of possible signals from light DM at CDMS and other experiments will be confirmed by new data.

In general, there is an important interplay between DM and the Higgs sector through the scattering WIMP cross section, the neutralino 
relic density and the invisible Higgs width, which needs to be systematically investigated in the coming years, as new results from DM direct 
detection and the LHC experiments will become available. 

%

\begin{acknowledgments}

We would like to thank B.~Allanach for useful discussion, B.~O'Leary for introducing us to the {\tt Vevacious} 
program, B.~Fuks for his advices on mono-jet cross section calculations, F.~Boudjema and 
A.~Pukhov for clarifications about the scattering cross section calculations in {\tt micrOMEGAs}. 

\end{acknowledgments}


\end{document}